\documentclass[12pt,a4paper]{article}
\pdfoutput=1

\usepackage[margin=1in]{geometry}
\usepackage{times}

\usepackage{color}
\usepackage[sort&compress,super,comma]{natbib}
\usepackage{amsmath,amsthm,amssymb,bm}
\usepackage{graphicx}
\usepackage{setspace}

\newcommand\bP{{\mathbf{P}}}
\newcommand\bW{{\mathbf{W}}}

\newcommand\bV{{\mathbf{V}}}

\newcommand\um{{\mu\text{m}}}
\newcommand\bN{{\mathbf{N}}}

\begin{document}

\onehalfspacing

\begin{center}{
\Large\bf
Ferromagnetic and antiferromagnetic order\\in bacterial vortex lattices\footnote{Originally submitted version; revised manuscript in press at {\itshape Nature Physics}.}
}
\end{center}

\singlespacing
\vspace{-1.0cm}

\begin{center}
\large
\noindent
Hugo Wioland$^{1,2^\dag}$, Francis G. Woodhouse$^{1^\dag,3}$, J\"orn Dunkel$^4$,\\and Raymond E. Goldstein$^{1,\ast}$
\end{center}

\noindent
$^1$Department of Applied Mathematics and Theoretical Physics, University of Cambridge, Wil\-berforce Road, Cambridge CB3 0WB, U.K. \\[0.1cm]
$^2$Institut Jacques Monod, Centre Nationale pour la Recherche Scientifique (CNRS), UMR 7592, Universit\'e Paris Diderot, Sorbonne Paris Cit\'e, F-75205 Paris, France \\[0.1cm]
$^3$Faculty of Engineering, Computing and Mathematics, The University of Western Australia, 35 Stirling Highway, Crawley, Perth WA 6009, Australia \\[0.1cm]
$^4$Department of Mathematics, Massachusetts Institute of Technology, 77 Massachusetts Avenue, Cambridge MA 02139-4307, U.S.A. \\[0.1cm]
{$^\dag${\it Present address.} $^\ast${\it Corresponding author.}}

\vspace{0.1cm}

\doublespacing

\noindent
\textbf{Despite their inherent non-equilibrium nature~\cite{1944Sc}, living systems can self-organize in highly ordered collective states\cite{2012Vicsek,2013Marchetti_Review} that share striking similarities with the thermodynamic equilibrium phases~\cite{2007Kardar,1979Me} of conventional condensed matter and fluid systems.  Examples range from the liquid-crystal-like arrangements of bacterial colonies~\cite{2004BenJacob,2008Volfson_PNAS}, microbial suspensions~\cite{2005Riedel_Science,2013Dunkel_PRL}  and tissues~\cite{Wu:2013aa} to the coherent macro-scale dynamics in schools of fish~\cite{2011Couzin} and flocks of birds~\cite{2009Parisi}. Yet, the generic mathematical principles that 
govern the emergence of structure in such artificial~\cite{2012Sanchez_Nature} and biological~\cite{2005Riedel_Science,2004BenJacob,2008Volfson_PNAS,2013Dunkel_PRL,2012Sokolov} systems are elusive. It is not clear when, or even whether, well-established theoretical concepts describing universal thermostatistics of 
equilibrium systems can capture and classify ordered states of living matter.
Here, we connect these two previously disparate regimes:
Through microfluidic experiments and mathematical modelling, we demonstrate that lattices of 
hydrodynamically coupled bacterial vortices can spontaneously organize into distinct phases of ferro- and antiferromagnetic order. 
The preferred phase can be controlled by tuning the vortex coupling through changes of the inter-cavity gap widths. The emergence of opposing order regimes is tightly linked to the existence of geometry-induced edge currents~\cite{2013Wioland_PRL,2014Lushi_PNAS}, reminiscent of those in quantum systems~\cite{PhysRevB.38.9375,PhysRevLett.95.226801,RevModPhys.81.109}.  Our experimental observations can be rationalized in terms of a generic lattice field theory, suggesting that bacterial spin networks belong to the same universality class as a wide range of equilibrium systems.  }

\onehalfspacing
\par

Lattice field theories (LFTs) have been instrumental in uncovering a wide range of fundamental physical phenomena, from quark confinement in atomic nuclei~\cite{1974Wilson} and neutron stars~\cite{2000Glendenning} to topologically protected states of matter~\cite{PhysRevC.73.055205} and transport in novel magnetic\cite{Nagaosa:2013aa} and electronic~\cite{Novoselov:2005aa,2009Graphene} materials.  LFTs can be constructed either by discretizing the space-time~\cite{1974Wilson}  continuum underlying classical and quantum field theories, or by approximating  discrete physical quantities, such as the electron spins in a crystal lattice, through continuous variables. In equilibrium thermodynamics, LFT approaches have proved invaluable both computationally and analytically, for a single LFT often represents a broad class of microscopically distinct physical systems that exhibit the same universal scaling behaviours in the vicinity of a phase transition~\cite{2007Kardar,1992Fernandez_Book}. However, until now there has been little evidence as to whether the emergence of order in living matter can be understood within this universality framework. Our combined experimental and theoretical analysis reveals a number of striking analogies between the collective cell dynamics in bacterial fluids and known phases of condensed matter systems, thereby implying that universality concepts may be more broadly applicable than previously thought.

\par

To realize a microbial non-equilibrium LFT, we injected dense suspensions of the rod-like swimming bacterium \textit{Bacillus subtilis} into shallow polydimethyl siloxane (PDMS) chambers in which identical circular cavities are connected to form one- and two-dimensional (2D) lattice networks  (Fig.~1 and Extended Data Fig.~5; Methods). Each cavity is $50\,\um$ in diameter and $18\,\um$ deep, a geometry known to induce a stably circulating vortex when a dense bacterial suspension is confined within an isolated flattened droplet~\cite{2013Wioland_PRL}.  For each cavity $i$, we define the continuous vortex spin variable  $V_i(t)$ at time~$t$ as the total angular momentum of the local bacterial flow within this cavity, determined by particle imaging velocimetry (PIV) analysis (Fig.~1b,f; Supplementary Videos~1 \& 2; Methods). To account for the effect of  oxygenation variability on suspension motility~\cite{2013Dunkel_PRL}, flow velocities are normalized by the overall root mean square (RMS) speed measured in the corresponding experiment.  Bacterial vortices in neighbouring cavities $i\sim j$ interact through a gap of predetermined width~$w$ (Fig.~1f). To explore different interaction strengths, we performed experiments over a range of gap parameters~$w$ (Methods). For square lattices, we varied $w$ from 4 to 25$\,\mu$m and found that for all but the largest gaps, $w\le w_*\approx 20\,\mu$m, the suspensions generally self-organize into coherent vortex lattices, exhibiting extended  domains of well-defined magnetic order whose characteristics depend on coupling strength (Fig.~1a,e).  If the gap size exceeds~$w_*$,  bacteria can move freely between cavities and individual vortices cease to exist. {A similar order--disorder transition is seen in triangular lattices (Fig.~3). Here, we focus exclusively on the vortex regime~$w< w_*$ and quantify magnetic order  through the normalized mean spin-spin correlation $\chi = \langle V_i(t) V_j(t) : i\sim j \rangle / \langle |V_i(t) V_j(t)| : i\sim j \rangle$, where $\langle \, \cdot \, : i\sim j \rangle$ denotes an average over time  and pairs $\{i,j\}$ of adjacent cavities (Methods).

\par

Square lattices reveal two distinct states of global magnetic order~(Fig.~1a,e,i), one with $\chi<0$ and the other with $\chi>0$, transitioning between them at a critical gap width~$w_\text{crit} \approx 8\,\mu\text{m}$ (Fig.~1j). For subcritical values $w<w_\text{crit}$, we observe an antiferromagnetic phase with anti-correlated ($\chi<0$) spin orientations   between neighbouring chambers (Fig.~1a; Supplementary Video 1).  By contrast, for  $w>w_\text{crit}$, positively-correlated ($\chi>0$) domains of ferromagnetic order are predominant (Fig.~1e; Supplementary Video 2).  Noting that the RMS spin $\langle V_i(t)^2 \rangle^{1/2}$ decays only slowly with increasing gap width $w\to w_*$~(Fig.~1k), and that the chambers do not impose any preferred handedness on the vortex spins (Extended Data Fig.~1),  we conclude that the observed phase behaviour is caused by spin-spin interactions. This is supported by the observation that although in both phases the individual spin amplitudes and orientations  fluctuate randomly over time, the correlation between two neighbouring vortex spins is typically conserved (Fig.~1i). Thus, although the bacterial vortex spins~$\{V_i(t)\}$ define a real-valued lattice field, the phenomenology of these continuous bacterial spin lattices is qualitatively similar to that of the classical 2D Ising model~\cite{2007Kardar}  with discrete binary spin variables $s_i\in\{\pm 1\}$, whose configurational probability  at finite temperature  $T=(k_B\beta)^{-1}$} is described by a thermal Boltzmann distribution $\propto\exp(-\beta J\sum_{i\sim j}s_is_j)$, where $J>0$ corresponds to ferromagnetic  and $J<0$ to antiferromagnetic order. The detailed theoretical analysis below shows that the observed phases in the bacterial spin system can be understood quantitatively in terms of a generic quartic LFT comprising two dual interacting lattices. The introduction of a double lattice is necessitated by the microscopic structure of the underlying bacterial flows.  By analogy with a lattice of interlocking cogs, one might have intuitively expected that the antiferromagnetic phase would be favoured, since only in this configuration does the bacterial flow along the cavity boundaries conform across the inter-cavity gap, thereby minimizing viscous stresses in the fluid~(Fig.~1b,c).  However, the extent of the observed ferromagnetic phase highlights a competing biofluid-mechanical effect.

 \par

Just as the quantum Hall effect~\cite{PhysRevB.38.9375} and the transport properties of graphene~\cite{PhysRevLett.95.226801,RevModPhys.81.109} arise from electric edge currents, the opposing order regimes observed here are explained by the existence of analogous bacterial edge currents.
At the boundary of an isolated flattened droplet of a bacterial suspension, a single layer of cells---an edge current---can be observed swimming against the bulk circulation~\cite{2013Wioland_PRL}. This narrow cell layer is key to the suspension dynamics: the hydrodynamics of the edge current circulating in one direction advects nearby cells in the opposite direction, which in turn dictate the bulk circulation by flow continuity~\cite{2014Lushi_PNAS}. Identical edge currents are present in our lattices (Supplementary Video 3) and explain both order regimes as follows. In the antiferromagnetic regime, when $w < w_\text{crit}$, the edge current driving a particular vortex will pass over the gap without leaving the cavity (Fig.~1c). Interaction with a neighbouring edge current through the gap favours parallel flow, inducing counter-circulation of neighbouring vortices and therefore driving antiferromagnetic order (Fig.~1d). However, when $w > w_\text{crit}$, the edge currents can no longer pass over the gaps and instead wind around the star-shaped pillars dividing the cavities (Fig.~1g). A clockwise (resp.\ counter-clockwise) edge current on a pillar induces counter-clockwise (resp.\ clockwise) circulation about the pillar in a thin region near its boundary. Flow continuity then induces clockwise (resp.\ counter-clockwise) flow in all cavities adjacent to the pillar, resulting in ferromagnetic order (Fig.~1h).
Thus by viewing the system as an anti-cooperative Union Jack lattice~\cite{1966Vaks_JETPUSSR,1970Stephenson_PRB} of both bulk vortex spins $V_{i}$ and near-pillar circulations $P_{j}$, we accommodate both order regimes: antiferromagnetism as indefinite circulations $P_j = 0$ and alternating spins $V_i = \pm V$ (Fig.~1d), and ferromagnetism as definite circulations $P_j = -P$ and uniform spins $V_i = V$ (Fig.~1h). To verify these considerations, we determined the net near-pillar circulation $P_j(t)$ using PIV (Methods)  and found that the RMS circulation $\langle P_j(t)^2 \rangle^{1/2}$ shows the expected monotonic increase as the inter-cavity gap widens~(Fig.~1k).

\par

Competition between the vortex--vortex and vortex--pillar interactions determines the resultant order regime. Their relative strengths can be inferred by mapping each experiment onto a continuous-spin Union Jack lattice (Fig.~1d,h). In this model, the interaction energy of the time-dependent vortex spins $\bV=\{V_{i}\}$ and pillar circulations $\bP=\{P_{j}\}$ is defined by the LFT  Hamiltonian
\begin{align}
\begin{split}
H(\bV,\bP) &= -J_v \sum_{V_i\sim V_j} V_i V_j
- J_p \sum_{V_i\sim P_j} V_i P_j
+ \sum_{V_i} \left( \tfrac{1}{2} a_v V_i^2 + \tfrac{1}{4} b_v V_i^4 \right) + \sum_{P_j} \tfrac{1}{2} a_p P_j^2.
\end{split}
\label{eq:full_Hamiltonian}
\end{align}
The first two sums are vortex-vortex and vortex-pillar interactions with strengths $J_v, J_p < 0$, where $\sim$ denotes adjacent lattice pairs. The last two sums are individual vortex and pillar circulation potentials.  Vortices must be subject to a quartic potential function with $b_v > 0$ to allow for a potentially double-welled potential if $a_v < 0$, encoding the observed symmetry breaking into spontaneous circulation absent other interactions~\cite{2012Woodhouse_PRL,2013Wioland_PRL}. In contrast, our data analysis implies that pillar circulations are sufficiently described by a quadratic potential of strength $a_p > 0$~(Extended Data Fig.~3; Methods).  To account for the experimentally observed spin fluctuations (Fig.~1i), we model  the dynamics of the lattice fields $\bV$ and $\bP$ through the coupled stochastic differential equations (SDEs)
\begin{subequations}
\label{e:SDE_model}
\begin{align}
d\bV &= -(\partial H/\partial \bV)dt + \sqrt{2 T_v} d\bW_v,\\
d\bP &= -(\partial H/\partial \bP)dt + \sqrt{2 T_p} d\bW_p,
\end{align}
\end{subequations}
where $\bW_v$ and $\bW_p$ are vectors of uncorrelated Wiener processes representing intrinsic and thermal fluctuations. The parameters $T_v$ and $T_p$ set the strength of random perturbations from energy-minimizing behaviour. In the limit case $T_v = T_p = T$, the stationary statistics of the solutions of Eq.~\eqref{e:SDE_model} is given by the Boltzmann distribution $\propto e^{-H/T}$. We inferred all seven parameters of the full SDE model for each experiment by linear regression on a  discretization of the SDEs~(Extended Data Fig.~2; Methods). As a cross-validation, we fitted appropriate functions of gap width $w$ to these estimates and simulated the resulting SDE model over a range of $w$ on a $6 \times 6$ lattice concordant with the observations (Methods). The agreement between experimental data and the numerically obtained vortex--vortex correlation $\chi(w)$ supports the validity of the double-lattice model (Fig.~1j).

\par

To reconnect with the classical 2D Ising model and understand better the experimentally observed phase transition, we project the Hamiltonian \eqref{eq:full_Hamiltonian} onto an effective square lattice model by making a mean-field assumption for the pillar circulations. In the experiments, $P_i$ is linearly correlated with the average spin of its vortex neighbours $[ P_i ]_V = \tfrac{1}{4}\sum_{j\, :\,  V_j \sim P_i} V_j$, with a constant of proportionality $-\alpha < 0$ only weakly dependent on gap width (Extended Data Fig.~3; Methods). Replacing $P_i \rightarrow -\alpha[P_i]_V$ as a mean field variable in the model eliminates all pillar circulations, yielding a standard quartic LFT for~$\bV$ (Methods). The mean-field dynamics are governed by the effective SDE $d\bV = -(\partial \hat H/\partial \bV)dt + \sqrt{2T} d\bW$ with energy
\begin{align*}
\hat H(\bV) =   -J \sum_{V_i \sim V_j} V_i V_j + \sum_{V_i} \left( \tfrac{1}{2} a V_i^2 + \tfrac{1}{4} b V_i^4 \right),
\end{align*}
which has steady-state probability density $p(\bV) \propto e^{-\beta \hat H}$ with $\beta = 1/T$. Note that in the limit $a \rightarrow -\infty$ and $b \rightarrow +\infty$ with $a/b$ fixed, the classical two-state Ising model is recovered by identifying $s_i=V_i /\sqrt{|a|/b} \in \{ \pm 1\}$. The reduced coupling constant $J$ relates to those of the double-lattice model $(J_v,J_p)$ in the thermodynamic limit as $J \approx J_v - \tfrac{1}{2}\alpha J_p$ (Methods), making manifest how competition between $J_v$ and $J_p$ can result in both antiferromagnetic ($|J_v|>\tfrac{1}{2}\alpha |J_p|$) or ferromagnetic ($|J_v|<\tfrac{1}{2}\alpha |J_p|$) behaviour.
We estimated $\beta J$, $\beta a$ and $\beta b$ for each experiment by directly fitting the effective one-spin potential $\mathcal{V}^\text{eff}(V \,|\, [V]_V) = -4 \beta J V [V]_V + \tfrac{1}{2}\beta aV^2 + \tfrac{1}{4}\beta bV^4$ via the log-likelihood $\log p(V  \,|\, [V]_V) = -\mathcal{V}^\text{eff} + \text{const}$ (Fig.~2; Extended Data Fig.~2; Methods). These estimates match those obtained independently using SDE regression methods (Fig.~2a--c; Methods), and show the transition from antiferromagnetic interaction ($\beta J < 0$) to ferromagnetic interaction ($\beta J > 0$) at $w_\text{crit}$~(Fig.~2a). As the gap width increases, the energy barrier to spin change falls (Fig.~2b) and the magnitude of the lowest energy spin decreases (Fig.~2c) due to weakening confinement within each cavity, visible as a flattening of the one-spin effective potential $\mathcal{V}^\text{eff}$  (Fig.~2d--f; Extended Data Fig.~2).

\par

Experiments on lattices of different symmetry groups lend further insight into the competition between edge currents and bulk flow. Unlike their square counterparts, triangular lattices cannot support antiferromagnetic states without frustration. Therefore, ferromagnetic order should be enhanced in a triangular bacterial spin lattice. This is indeed observed in our experiments: at moderate gap size $w\lesssim18\,\mu$m, we found  exclusively a highly robust ferromagnetic phase of either handedness (Fig.~3a,b,d; Supplementary Video 4), reminiscent of quantum vortex lattices in Bose-Einstein condensates~\cite{2001AboShaeer_Science}. At comparable gap size, the spin correlation is approximately 4 to 8 times larger than in the square lattice. Increasing the gap size beyond $20\,\mu$m eventually destroys the spontaneous circulation within the cavities and a disordered state prevails (Fig.~3c,d), with a sharper transition than for the square lattices (Fig.~1j). Conversely, a 1D line lattice exclusively exhibits antiferromagnetic order  as the suspension is unable to maintain the very long uniform edge currents that would be necessary to sustain a ferromagnetic state (Extended Data Fig. 5). These results manifest the importance of lattice geometry and dimensionality for vortex ordering in bacterial spin lattices, in close analogy with their electromagnetic counterparts.

\par
In conclusion,  understanding the ordering principles of microbial matter is a key challenge in active materials design~\cite{2012Sanchez_Nature},  quantitative biology and biomedical research. Improved prevention strategies for pathogenic biofilm formation, for example, will require detailed knowledge of how bacterial flows interact with complex porous surface structures. Our study shows that collective excitations in geometrically confined bacterial suspensions can spontaneously organize in phases of magnetic order that can be robustly controlled by edge currents.  These results demonstrate fundamental similarities with a broad class of widely studied quantum systems~\cite{PhysRevB.38.9375,2001AboShaeer_Science,RevModPhys.81.109}, suggesting that theoretical concepts originally developed to describe magnetism in disordered media could potentially capture microbial behaviours in complex environments.
Future studies may try to explore further the range and limits of this promising analogy.

%%%%%%%%%%%%%%%%%%%%%%%%%%%%%%%%%%%%%%%%%%%%%%%%%%%%%%%%%%%%%%
% ENDNOTES
%%%%%%%%%%%%%%%%%%%%%%%%%%%%%%%%%%%%%%%%%%%%%%%%%%%%%%%%%%%%%%

\small

\paragraph*{Acknowledgements}
We thank Vasily Kantsler and Enkeleida Lushi for assistance and discussions.
This work was supported by European Research Council Advanced Investigator Grant 247333  (H.W. and R.E.G.), EPSRC (H.W. and R.E.G.), an MIT Solomon Buchsbaum Award (J.D)   and an Alfred P. Sloan Research Fellowship (J.D.).

\paragraph*{Author contributions}
All authors designed the research and collaborated on theory. H.W. performed experiments and PIV analysis. H.W. and F.G.W. analysed PIV data and performed parameter inference. F.G.W. and J.D. wrote simulation code. All authors wrote the paper.

\normalsize

\clearpage

%%%%%%%%%%%%%%%%%%%%%%%%%%%%%%%%%%%%%%%%%%%%%%%%%%%%%%%%%%%%%%
% FIGURES
%%%%%%%%%%%%%%%%%%%%%%%%%%%%%%%%%%%%%%%%%%%%%%%%%%%%%%%%%%%%%%

%%%%%%%%%%%%%%%% FIG 1
\begin{figure*}[h!]
\centering
\includegraphics[width=\textwidth]{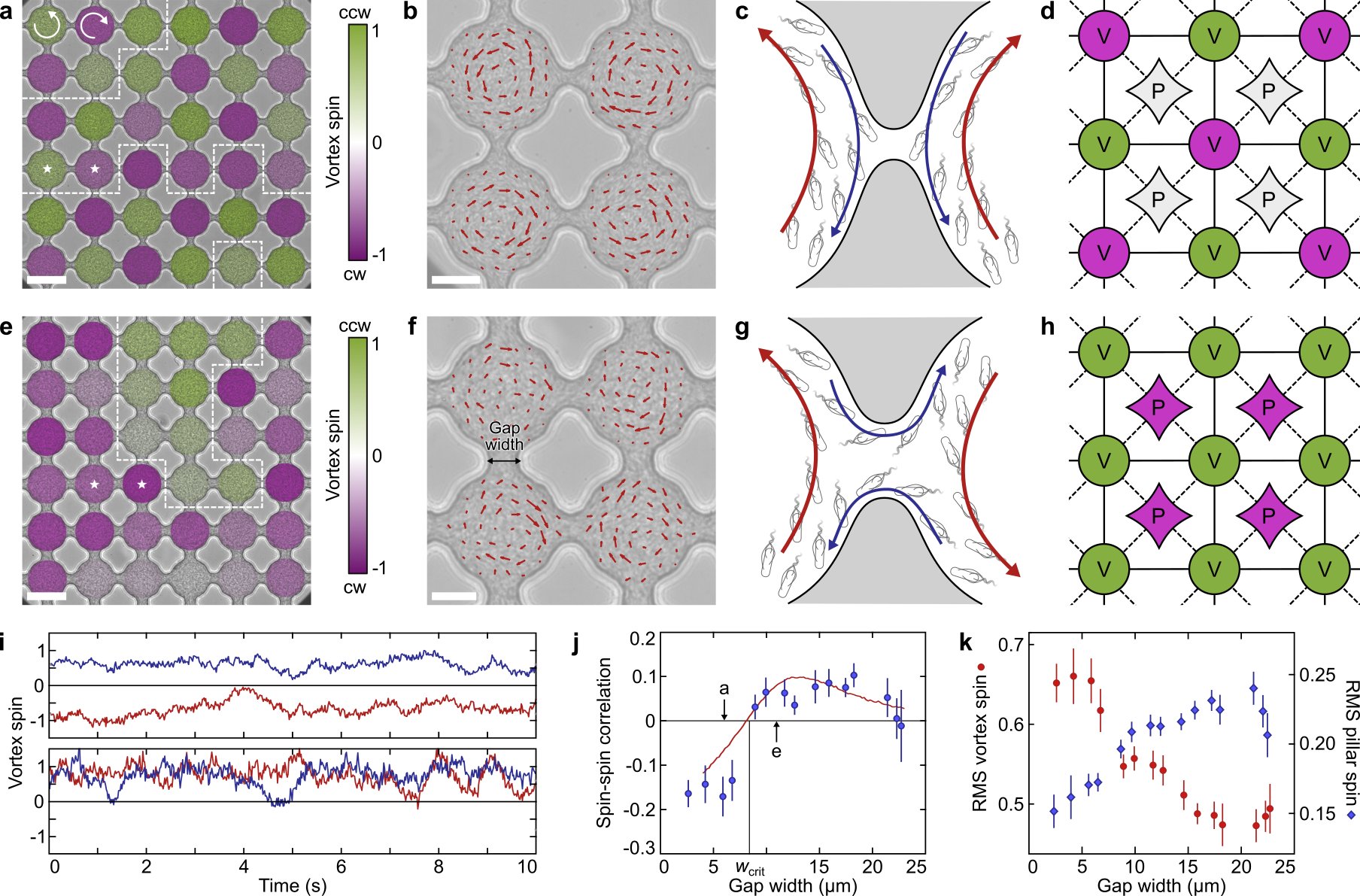}
\caption{
\onehalfspacing
\small
\textbf{Edge currents determine antiferromagnetic and ferromagnetic order in  a square lattice of bacterial vortices.} 
\textbf{a}, 
Three domains of antiferromagnetic order highlighted by dashed white lines (gap width $w=6\,\mu$m). Scale bar: $50\,\um$. Overlaid false colour shows spin magnitude (see Supplementary Video 1 for raw data).
\textbf{b}, 
Bacterial flow PIV field within an antiferromagnetic domain (Supplementary Video 1). For clarity, not all velocity vectors are shown. Largest arrows correspond to speed $40\,\mu$m/s. Scale bar: $20\,\um$.
\textbf{c}, 
Schematic of bacterial flow circulation in the vicinity of a gap.  For small gaps $w<w_\text{crit}$, bacteria forming the edge currents (blue arrows) swim across the gap, remaining in their original cavity. Bulk flow (red) is directed opposite to the edge current~\cite{2013Wioland_PRL,2014Lushi_PNAS}  (Supplementary Video 3).
\textbf{d}, 
Graph of the Union Jack double-lattice model in an antiferromagnetic state with zero net pillar circulation. 
Solid and dashed lines depict vortex--vortex and vortex--pillar interactions of respective strengths $J_v$ and $J_p$.
Vortices and pillars are colour-coded according to their spin. 
\textbf{e}, 
For supercritical gap widths $w>w_\text{crit}$,  extended domains of ferromagnetic order predominate (Supplementary Video 2; $w = 11\,\mu$m). Scale bar: $50\,\um$.
\textbf{f}, 
PIV field within a ferromagnetic domain (Supplementary Video 2). Largest arrows: $36\,\mu$m/s. Scale bar: $20\,\um$.
\textbf{g}, 
For $w>w_\text{crit}$, bacteria forming the edge current (blue arrows) swim along the PDMS boundary through the gap, driving bulk flows (red) in the opposite directions, thereby aligning neighbouring vortex spins.
\textbf{h}, 
Ferromagnetic state of the Union Jack lattice induced by edge current loops around the pillars. 
 \textbf{i}, 
Trajectories of neighbouring spins ($\star$-symbols in \textbf{a},\textbf{e}) fluctuate over time, signalling a non-zero effective temperature,  but their relative orientation remains conserved (top: antiferromagnetic; bottom: ferromagnetic).
\textbf{j}, 
The zero of the spin-spin correlation $\chi$ at $w_{\text{crit}}\approx8\,\um$ marks the phase transition. 
The best-fit Union Jack model (solid line) is consistent with the experimental data.
\textbf{k}, 
RMS vortex spin $\langle V_i^2 \rangle^{1/2}$ decreases with the gap size $w$, showing weakening of the circulation. RMS pillar spin $\langle P_j^2 \rangle^{1/2}$ increases with $w$, reflecting enhanced bacterial circulation around pillars.
Each point in \textbf{j},\textbf{k} represents an average over $\geq 5$ movies in $3\,\um$ bins at $1.5\,\um$ intervals; vertical bars indicate standard errors (Methods).
}
\end{figure*}

%%%%%%%%%%%%%%%% FIG 2
\begin{figure}[h!]
\centering
\includegraphics[width=120mm]{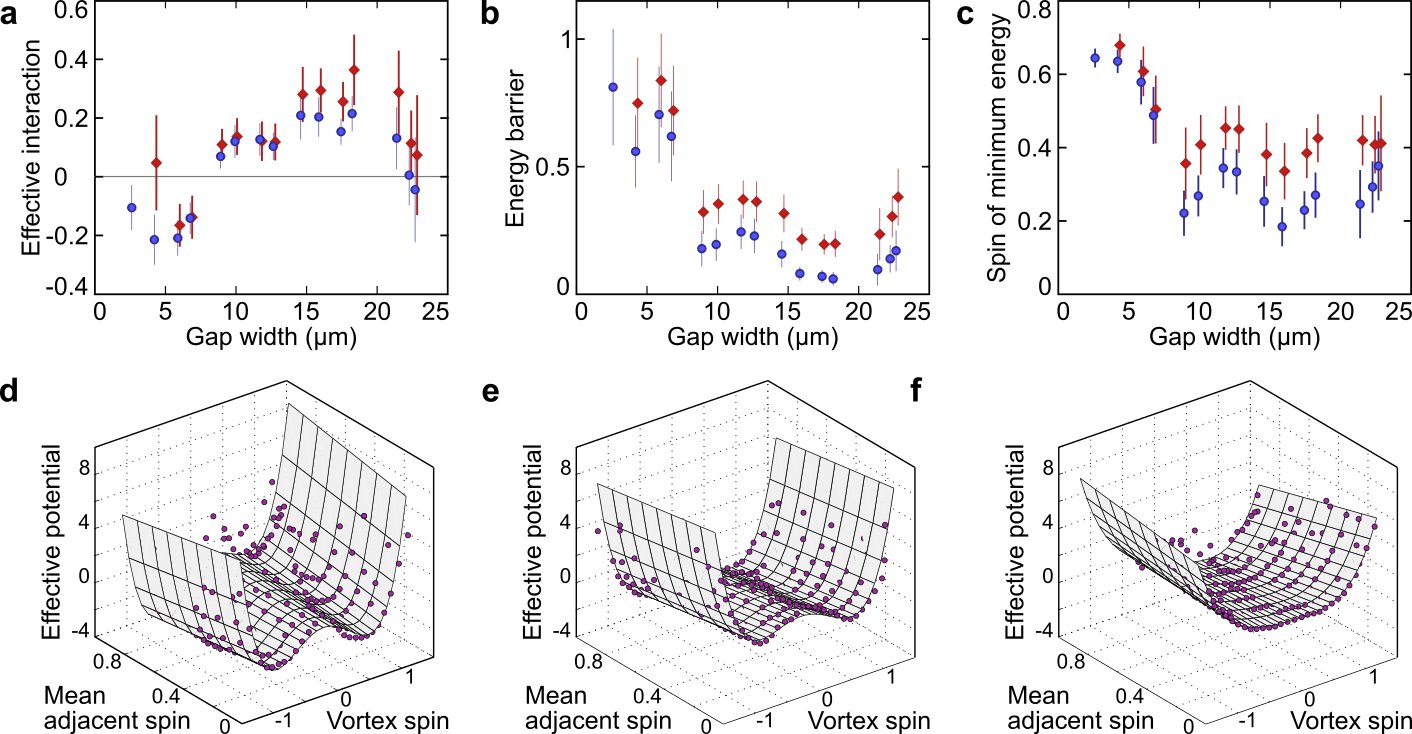}
\caption{
\onehalfspacing
\small
\textbf{Best-fit mean-field LFT model captures the phase transition in the square lattice.}
\textbf{a}, 
A sign change of the effective interaction $\beta J$ signals the transition from antiferro- to ferromagnetic states. 
\textbf{b}, 
The effective energy barrier ${\beta a^2}/(4b)$ (Methods) decreases with the gap size $w$, reflecting increased 
susceptibility to fluctuations.
\textbf{c}, 
The spin $V_\text{min}$ minimizing the single-spin potential  (Methods) decreases with $w$ in agreement with the decrease in the RMS vortex spin (Fig.~1f).
Each point in \textbf{a}--\textbf{c} represents an average over $\geq 5$ movies in $3\,\um$ bins at $1.5\,\um$ intervals; vertical bars indicate standard errors (Methods).
\textbf{d-f}, Examples of effective single-spin potential $\mathcal{V}^\text{eff}$ conditional on the mean spin of adjacent vortices $[V]_V$. Data (points) and estimated potential (surface) for three movies with gap widths $6$, $10$ and $17\,\um$.
}
\end{figure}

%%%%%%%%%%%%%%%% FIG 3
\begin{figure}[h!]
\centering
\includegraphics[width=89mm]{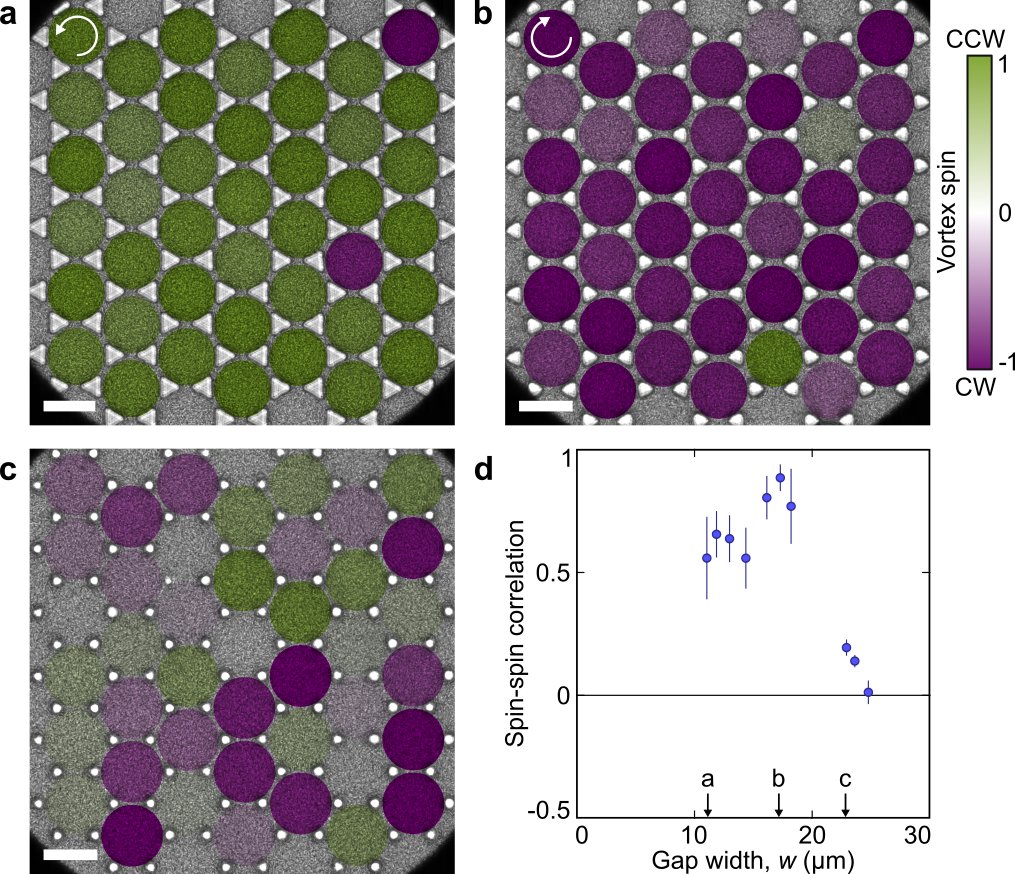}
\caption{
\onehalfspacing
\small
\textbf{Frustration in triangular lattices determines the preferred order.}
\textbf{a},\textbf{b}, 
Triangular lattices favour ferromagnetic states of either handedness  (Supplementary Video 4). Vortices are colour-coded by spin.
\textbf{c}, 
At the largest gap size, bacterial circulation becomes unstable. Scale bar: $50\,\um$.
\textbf{d}, 
The spin-spin correlation $\chi$ shows strongly enhanced ferromagnetic order compared with the square lattice (Fig.~1j).
Each point represents an average over $\geq 5$ movies in $3\,\um$ bins at $1.5\,\um$ intervals; vertical bars indicate standard errors (Methods).}
\end{figure}

\clearpage

%%%%%%%%%%%%%%%%%%%%%%%%%%%%%%%%%%%%%%%%%%%%%%%%%%%%%%%%%%%%%%
% MAIN TEXT REFERENCES

%%%%%%%%%%%%%%%%%%%%%%%%%%%%%%%%%%%%%%%%%%%%%%%%%%%%%%%%%%%%%%

\clearpage

%%%%%%%%%%%%%%%%%%%%%%%%%%%%%%%%%%%%%%%%%%%%%%%%%%%%%%%%%%%%%%
% METHODS
%%%%%%%%%%%%%%%%%%%%%%%%%%%%%%%%%%%%%%%%%%%%%%%%%%%%%%%%%%%%%%

\section*{Methods}

\subsubsection*{Experiments}
Wild-type \textit{Bacillus subtilis} cells (strain 168) were grown in Terrific Broth (Sigma). A monoclonal colony was transferred from an agar plate to $25\,$mL of medium and left to grow overnight at $35^\circ$C on a shaker. The culture was diluted $200$-fold into fresh medium and harvested after approximately $5$ hours when more than $90\%$ of the bacteria were swimming. $10\,$mL of the suspension was then concentrated by centrifugation at $1500g$ for $10\,$minutes, resulting in a pellet with volume fraction approximately $20$\% which was used without further dilution. 

The microchambers were made of polydimethyl siloxane (PDMS) bound to a glass coverslip by oxygen plasma etching. These comprised a square, triangular or linear lattice of $\sim\!18\,\um$-deep circular cavities with $60\,\um$ between centres, each of diameter $\sim\!50\,\um$, connected by $4$--$25\,\um$-wide gaps for linear and square lattices (Fig.~1a,e; Extended Data Fig.~5) and $10$--$25\,\um$-wide gaps for triangular lattices (Fig.~3a--c). The smallest possible gap size was limited by the fidelity of the etching.

Approximately $5\,\mu$L of the concentrated suspension was manually injected into the chamber using a syringe. Both inlets were then sealed to prevent external flow. 
We imaged the suspension on an inverted microscope (Zeiss, Axio Observer Z1) under bright field illumination, through a $40\times$ oil-immersion objective. Movies $10\,$s long were recorded at $60\,\text{f.p.s.}$ on a high-speed camera (Photron Fastcam SA3) at $4$ and $8$ minutes after injection.
Though the PDMS lattices were typically $\sim\!15$ cavities across, to avoid boundary effects we only imaged a central subregion spanning $6 \times 6$ cavities for square lattices, $7 \times 6$ cavities for triangular lattices, and $7$ cavities for linear lattices (multiple of which were captured on a single slide).

Fluorescence in Supplementary Video 3 was achieved by labelling the membranes of a cell subpopulation with fluorophore FM4-64 following the protocol of Lushi et al.\cite{2014Lushi_PNAS}. The suspension was injected into an identical triangular lattice as in the primary experiments and imaged at $5.6\,\text{f.p.s.}$  on a spinning-disc confocal microscope through a $63\times$ oil-immersion objective.

\subsubsection*{Analysis}

For each frame of each movie, the aggregate bacterial flow field $\mathbf{u}(x,y,t)$ was measured by standard particle image velocimetry (PIV) without time averaging, using a customized version of mPIV~\cite{MPIV}. PIV subwindows were $16\times 16$ pixels with 50\% overlap, yielding $\sim\!150$ vectors per cavity per frame.
Cavity regions were identified in each movie by manually placing the centre and radius of the bottom left cavity, measuring vectors to its immediate neighbours, and repeatedly translating to generate the full grid.
Pillar edges were then calculated from the cavity grid and the gap width (measured as the minimum distance between adjacent pillars).

The spin $V_i(t)$ of each circular cavity $i$ at time $t$ is defined as the normalized planar angular momentum
\begin{equation*}
	V_i(t) = \frac{\mathbf{\hat z} \cdot \left[\sum_{(x,y)_i} \mathbf{r}_i(x,y) \times \mathbf{u}(x,y,t) \right]}{\overline{U}\sum_{(x,y)_i} |\mathbf{r}_i(x,y)|},
\end{equation*} 
where $\mathbf{r}_i(x,y)$ is the vector from the cavity centre to $(x,y)$, and sums run over all PIV grid points $(x,y)_i$ inside cavity $i$. 
For each movie, we normalize velocities by the root-mean-square (RMS) suspension velocity $\overline{U} = \langle \mathbf{u}(x,y,t)^2 \rangle^{1/2}$, where the average is over all grid points $(x,y)$ and all times $t$, to account for the effects of variable oxygenation on motility~\cite{2013Dunkel_PRL}.
This definition has $V_i(t)>0$ for counter-clockwise spin and $V_i(t)<0$ for clockwise spin. A vortex of radially-independent speed, i.e.\ $\mathbf{u}(x,y,t) = u \bm{\hat \theta}$ where $\bm{\hat \theta}$ is the azimuthal unit vector, has $V_i(t) = \pm1$; conversely, randomly oriented flow has $V_i(t) = 0$.  The average spin--spin correlation $\chi$ of a movie is then defined as
\begin{equation*}
	\chi = \frac{\langle V_i(t) V_j(t) : i \sim j \rangle}{\langle |V_i(t) V_j(t)| : i \sim j \rangle},
\end{equation*}
where $\langle \cdot : i \sim j \rangle$ denotes an average over all frames and all adjacent pairs of vortices $\{ i,j \}$. If all vortices share the same sign, then $\chi=1$ (ferromagnetism);  if each vortex is of opposite sign to its neighbours, then $\chi=-1$ (antiferromagnetism); if the vortices are uniformly random, then $\chi=0$.

Similarly, the circulation $P_j(t)$ about pillar $j$ at time $t$ is defined as the normalized average tangential velocity
\begin{equation*}
P_j(t) = \frac{\sum_{(x,y)_j} \mathbf{u}(x,y,t)\cdot \mathbf{\hat t}_j(x,y)}{\overline{U}\sum_{(x,y)_j} 1},
\end{equation*} 
where $\mathbf{\hat t}_j(x,y)$ is the unit vector tangential to the pillar, and sums run over PIV grid points $(x,y)_j$ closer than $5\,\um$ to the pillar $j$.

Results presented are typically averaged in bins of fixed gap width. All plots with error bars use $3\,\um$ large bins, calculated every $1.5\,\um$ ($50\%$ overlap), and bins with fewer than 5 movies were excluded. Error bars denote standard error. Bin counts for square lattices (Fig.~1j,k; Fig.~2a--c; Extended Data Fig.~3d) are 8, 8, 13, 14, 21, 27, 27, 22, 18, 22, 20, 11, 7, 13, 7; bin counts for triangular lattices (Fig.~3d) are 5, 14, 16, 13, 16, 15, 5, 5, 10, 7; and bin counts for linear lattices (Extended Data Fig.~5d) are 5, 7, 8, 8, 9, 9, 6, 5, 6, 7, 6, 8, 9, 5, 6, 5, 6.

\subsubsection*{Parameter inference under the full model}

For a given sequence of discrete experimental observations $\{\bV(t),\bP(t)\}_{t=n\Delta t}$ derived from one movie, with constant time step $\Delta t = 1/60\,\text{s}$, we wish to estimate the most likely parameter values assuming the SDE model \eqref{e:SDE_model} holds. We do this by first discretizing Eq.~\eqref{e:SDE_model} and then applying linear regression.
First, the rescaling by $\overline{U}$ used to eliminate variable oxygenation effects implies that we must also rescale the time step to $\delta t = (\overline{U}/1\um)\times \Delta t$.
Now, using this time step, Eq.~\eqref{e:SDE_model}  discretizes in the Euler--Maruyama scheme~\cite{Higham2001_SIAMrev} as
\begin{align}
\bV(t+\delta t) &= \bV(t)  -(\partial H/\partial \bV)_t \delta t + \sqrt{2 T_v \delta t} \bN_v, \label{eq:full_model_disc_V} \\
\bP(t+\delta t) &= \bP(t)  -(\partial H/\partial \bP)_t \delta t + \sqrt{2 T_p \delta t} \bN_p, \label{eq:full_model_disc_E}
\end{align}
where $\bN_v$ and $\bN_p$ are vectors of independent $\mathcal{N}(0,1)$ random variables. Component-wise, Eqs.~\eqref{eq:full_model_disc_V} and \eqref{eq:full_model_disc_E} read
\begin{align}
\begin{split}
 V_i(t+\delta t) &=  (1 - a_v \delta t)V_i(t) -  b_v \delta t V_i(t)^3 \\ &\qquad\qquad + J_v \delta t \! \sum_{j \,:\,  V_{j} \sim V_{i}} \! V_j(t)
 + J_p \delta t \! \sum_{j \,:\,  P_{j} \sim V_{i}} \! P_j(t)  + \sqrt{2 T_v \delta t} N_{v,i}, 
 \end{split}
\label{eq:full_model_reg_V}\\
\label{eq:full_model_reg_E}
 P_i(t+\delta t) &=  (1 - a_p \delta t)P_i(t) + J_p \delta t \! \sum_{j \,:\, V_{j} \sim P_{i}} \! V_j(t)   + \sqrt{2 T_p \delta t} N_{p,i}.
\end{align}
By Eq.~\eqref{eq:full_model_reg_E}, using data from all observation times and vortices to perform a linear regression of $P_i(t+\delta t)$ on the two variables
\begin{align*}
\left\{ P_i(t), \sum_{ j\,:\, V_{j} \sim P_{i}} V_j(t) \right\}
\end{align*}
yields estimates $\{1 - \hat a_p \delta t, \hat J_p \delta t\}$ of the variables' respective coefficients and thence estimates $\hat a_p$ and $\hat J_p$ of $a_p$ and $J_p$.
Next, after substituting the estimate $J_p = \hat J_p$ into Eq.~\eqref{eq:full_model_reg_V} to reduce the dimensionality, a linear regression of $V_i(t+\delta t) - \hat J_p \delta t \sum_{ j\,:\, P_{j} \sim V_{i}} P_j(t) $ on the three variables
\begin{align*}
\left\{ V_i(t), V_i(t)^3, \sum_{j\,:\, V_j \sim V_i} V_j(t) \right\}
\end{align*}
yields estimates $\{1 - \hat a_v \delta t, -\hat b_v \delta t, \hat J_v \delta t\}$ of their respective coefficients and thence estimates $\hat a_v$, $\hat b_v$ and $\hat J_v$ of $a_v$, $b_v$ and $J_v$.
Finally,  the variances $2 T_v \delta t$ and $2 T_p \delta t$ of the residuals to the regressions in Eqs.~\eqref{eq:full_model_reg_V} and \eqref{eq:full_model_reg_E} respectively yield estimates $\hat T_v$ and $\hat T_p$ of $T_v$ and $T_p$.

\subsubsection*{Vortex--vortex correlation reconstruction}

To reconstruct the vortex--vortex correlation function $\chi(w)$ as a continuous function of gap width $w$, we first reconstructed the parameters as functions of $w$ from the experimental data and then simulated the model \eqref{e:SDE_model}  over a range of $w$, as follows.
Running the parameter estimation procedure for every suitable experimental movie (those not containing any `locked' immobile cavities, occasionally seen at small $w$) results in a set of parameter estimates $\mathcal{E}_i$ at gaps $w_i$.
Estimates for movies from the same experiment were averaged, and then placed into non-overlapping $w$-bins of size $2.5\,\um$ and averaged in both $w$ and parameter value within each bin (Extended Data Fig.~2, points). Using non-linear least-squares regression estimation, the parameters were then fitted with chosen functional forms: $J_v,J_p,a_p$ with a logistic function $\alpha_1 + \alpha_2/(1 + 10^{\alpha_3 (\alpha_4 - w)})$; $a_v,b_v$ with a rational function $\alpha_1/(w + \alpha_2)$; and $T_v,T_p$ with a rational function $(\alpha_1 + \alpha_2 w)/(w^2 + \alpha_3 w + \alpha_4)$ (Extended Data Fig.~2, lines). These forms were chosen as appearing to give the best representation of the data points' behaviour (such as not introducing maxima where none are observed for $J_v, J_p, a_p$, and not presuming too detailed a functional form for the noisiest parameters $a_v$ and $b_v$) with the fewest possible fit parameters.
Finally, we numerically integrated Eq.~\eqref{e:SDE_model}  using the discretization in Eqs.~\eqref{eq:full_model_disc_V} and \eqref{eq:full_model_disc_E}, wherein we set $N=6$ and $\delta t = 1/600$. We initialized $\bV$ and $\bP$ to zero, and after a burn-in period of $50/\delta t$ frames we recorded every frame. Trial and error showed an observation period of $8000/\delta t$ to be sufficient to obtain a stable estimate of the average vortex--vortex correlation $\chi$, and this was further averaged over 25  repetitions at each of 101 regularly-spaced values of $w$ in the range $\min w_i \leq w \leq \max w_i$ (Fig.~1j).

\subsubsection*{Reduction to vortex-only model}

Integrating Eq.~\eqref{e:SDE_model}  shows that each pillar circulation $P_i$ sits in an effective potential
\begin{align*}
\mathcal{P}^\text{eff}(P_i | [P_i]_V) =  - 4 J_p P_i [P_i]_V + \tfrac{1}{2}a_p P_i^2,
\end{align*}
dependent only on the mean spin of adjacent vortices $[P_i]_V = \frac{1}{4} \sum_{j\,:\, V_j \sim P_i} V_j$. This suggests a mean-field reduction is possible if $[P_i]_V$ can be approximated in terms of $P_i$.
We found $P_i$ to be linearly correlated with $[P_i]_V$ in every square-lattice experiment (Extended Data Fig.\ 3a--c); writing $-\alpha$ for the correlation coefficient, we found $\alpha\approx 0.5$ with weak dependence on the gap width $w$ (Extended Data Fig.\ 3d).
We use this to reduce the dimensionality of the full model by approximating $P_i$ with $-\alpha [P_i]_V$. The vortex--pillar interaction becomes
\begin{align*}
\sum_{V_i \sim P_j} V_i P_j
= -\frac{\alpha}{2} \sum_{V_i\sim V_j}V_i V_j - \alpha \sum_{V_i} V_i^2  \, \text{ (+ n.n.n.)}, 
\end{align*}
where `n.n.n.' denotes next-nearest-neighbour interactions which we neglect. 
Therefore, after also neglecting independent fluctuations in $\bP$ (effectively setting $a_p \rightarrow 0$), the Hamiltonian \eqref{eq:full_Hamiltonian} reduces to
\begin{align*}
\hat H(\bV) =   -J \sum_{V_i \sim V_j} V_i V_j 
				+ \sum_{V_i} \left( \tfrac{1}{2} a V_i^2 + \tfrac{1}{4} b V_i^4 \right)
\end{align*}
where $J = J_v-\tfrac{1}{2}\alpha J_p$, $a = a_v + 2\alpha J_p$ and $b = b_v$.
Note these exact relations will only be achieved in the thermodynamic limit when boundary effects are eliminated.

\subsubsection*{Parameter inference under the reduced model by distribution fitting}

We assume that $\bV$ obeys a Boltzmann distribution $p(\bV) \propto e^{-\beta \hat{H}(\bV)}$.
The probability density $p(V_i \,|\, [V_i]_V)$ of one spin $V_i$ conditional on the mean of its adjacent spins $[V_i]_V = \tfrac{1}{4}\sum_{j\,:\, V_j\sim V_i}V_j$ satisfies
\begin{align}
\label{eq:M_Veff_p}
\log p(V_i \,|\, [V_i]_V) - \log p(0 \,|\, [V_i]_V)
						& = -\mathcal{V}^\text{eff}(V_i \,|\, [V_i]_V),
\end{align}
where we have defined the effective single-vortex potential
\begin{align*}
\mathcal{V}^\text{eff}(V_i \,|\, [V_i]_V) 
				& = - 4\beta J V_i[V_i]_V
					+ \tfrac{1}{2}\beta a V_i^2 
					+ \tfrac{1}{4}\beta b V_i^4.
\end{align*}
We estimate $p(V_i \,|\, [V_i]_V)$ for each movie by forming a two-dimensional histogram in $V_i$ and $[V_i]_V$ and then normalizing at every fixed $V_i$. In forming the histogram, we exploit the invariance of $\mathcal{V}^\text{eff}$ under the transformation $V_i\rightarrow -V_i$ and $[V_i]_V\rightarrow -[V_i]_V$ to double the number of data points.

Taking the antisymmetric part $\mathcal{V}^\text{eff}_\text{anti}
		 = \tfrac{1}{2} \left[ \mathcal{V}^\text{eff}(V_i \,|\, [V_i]_V)-\mathcal{V}^\text{eff}(-V_i \,|\, [V_i]_V) \right]$ eliminates the non-interacting terms, so Eq.~\eqref{eq:M_Veff_p} implies
\begin{align*}
 \tfrac{1}{2} \left[ - \log p(V_i\,|\,[V_i]_V) + \log p(-V_i\,|\,[V_i]_V) \right] 
		 = - 4\beta J V_i [V_i]_V.
\end{align*}
This allows estimation of the interaction constant $\beta J$ from the density $p(V_i | [V_i]_V)$ (Fig.\ 2a and Extended Data Fig.\ 4a--c).
The symmetric part $\mathcal{V}^\text{eff}_\text{sym}  = \tfrac{1}{2} \left[ \mathcal{V}^\text{eff}(V_i \,|\, [V_i]_V) + \mathcal{V}^\text{eff}(-V_i \,|\, [V_i]_V) \right]$ eliminates the interaction term in a similar fashion, so Eq.~\eqref{eq:M_Veff_p} now implies
\begin{align*}
  \frac{1}{2} \left[ - \log p(V_i\,|\,[V_i]_V) - \log p(-V_i\,|\,[V_i]_V) \right] + \log p(0\,|\,[V_i]_V)
		 = \frac{1}{2}\beta a V_i^2 + \frac{1}{4}\beta b V_i^4.
\end{align*}
Typically there are fewer observations near $V_i=0$, so $p(0\,|\,[V_i]_V)$ can be difficult to infer directly. Instead, we adjust $\log p(0 \,|\, [V_i]_V)$ for each $[V_i]_V$ bin to minimize the difference between $\mathcal{V}^\text{eff}_\text{sym}(V \,|\, [V_i]_V)$ and $\mathcal{V}^\text{eff}_\text{sym}(V \,|\, 0)$. We then fit the remaining single vortex potential with parameters $\beta a$ and $\beta b$ (Extended Data Fig.\ 4d--f), from which we compute the spins $\pm V_\text{min}$ ($V_\text{min} > 0$) minimizing the local effective single-spin energy $\mathcal{V}^\text{eff}_\text{sym}$, namely
\begin{align*}
V_\text{min} =
\begin{cases}
 \sqrt{|a|/b} & \text{if } a < 0, \\
 0 & \text{if } a > 0,
 \end{cases}
\end{align*}
and the effective spin-flip energy barrier
\begin{align*}
\mathcal{V}^\text{eff}_\text{sym}(0) - \mathcal{V}^\text{eff}_\text{sym}(\pm V_\text{min}) = 
\begin{cases}
\beta a^2/(4b) & \text{if } a < 0, \\
0 & \text{if } a > 0,
\end{cases}
\end{align*}
which together characterize the single-spin symmetric quartic potential (Fig.~2b,c).

\subsubsection*{Parameter inference under the reduced model by SDE discretization}

Estimations made by the above method were verified by estimations obtained through the same SDE discretization method used in the full model. The components of the reduced model SDE $d\bV = -(\partial \hat H/\partial \bV)dt + \sqrt{2T} d\bW$ have Euler--Maruyama discretization
\begin{align*}
 V_i(t+\delta t) &=  (1 - a \delta t)V_i(t) -  b \delta t V_i(t)^3
 + J \delta t \sum_{j \,:\,  V_{j} \sim V_{i}} V_j(t) + \sqrt{2T \delta t} N_{i},
\end{align*}
where $N_i$ are independent $\mathcal{N}(0,1)$ random variables.
Performing linear regression of ${V_i(t+\delta t)}$ on the three variables
\begin{align*}
\left\{ V_i(t), V_i(t)^3, \sum_{j\,:\, V_j \sim V_i} V_j(t) \right\}
\end{align*}
then gives estimates $\{1 - \hat a \delta t, -\hat b \delta t, \hat J \delta t\}$ of the respective coefficients, from which estimates $\hat a$, $\hat b$ and $\hat J$ of the variables $a$, $b$ and $J$ can be deduced. The estimate $\hat T$ of the fluctuation strength $T$ is estimated via the variance $2T\delta t$ of the residuals to the regression, which gives an estimate $\hat\beta = 1/\hat T$ of the inverse `temperature' $\beta$ in the Boltzmann distribution. The non-dimensional combinations $\hat\beta \hat J$, $\hat\beta \hat a$ and $\hat\beta \hat b$ can then be directly compared with the estimates obtained using the distribution-fitting method.

Though SDE discretization independently gives both temperature and coupling constants, it is likely to possess greater intrinsic bias than distribution fitting. Discretization was the only method open for the full model, as the SDE steady state cannot be solved analytically. However, since the reduced model allows for distribution fitting, coupling constant values obtained using that method are preferable, with SDE discretization functioning as an independent verification.

%%%%%%%%%%%%%%%%%%%%%%%%%%%%%%%%%%%%%%%%%%%%%%%%%%%%%%%%%%%%%%
% ADDITIONAL REFERENCES
%%%%%%%%%%%%%%%%%%%%%%%%%%%%%%%%%%%%%%%%%%%%%%%%%%%%%%%%%%%%%%

\renewcommand{\refname}{Additional References}

\clearpage

%%%%%%%%%%%%%%%%%%%%%%%%%%%%%%%%%%%%%%%%%%%%%%%%%%%%%%%%%%%%%%
% EXTENDED DATA FIGURES
%%%%%%%%%%%%%%%%%%%%%%%%%%%%%%%%%%%%%%%%%%%%%%%%%%%%%%%%%%%%%%

\section*{Extended Data Figures}

\setcounter{figure}{0}
\renewcommand{\figurename}{Extended Data Figure}

\begin{figure}[h!]
	\centering
	\includegraphics[width=89mm]{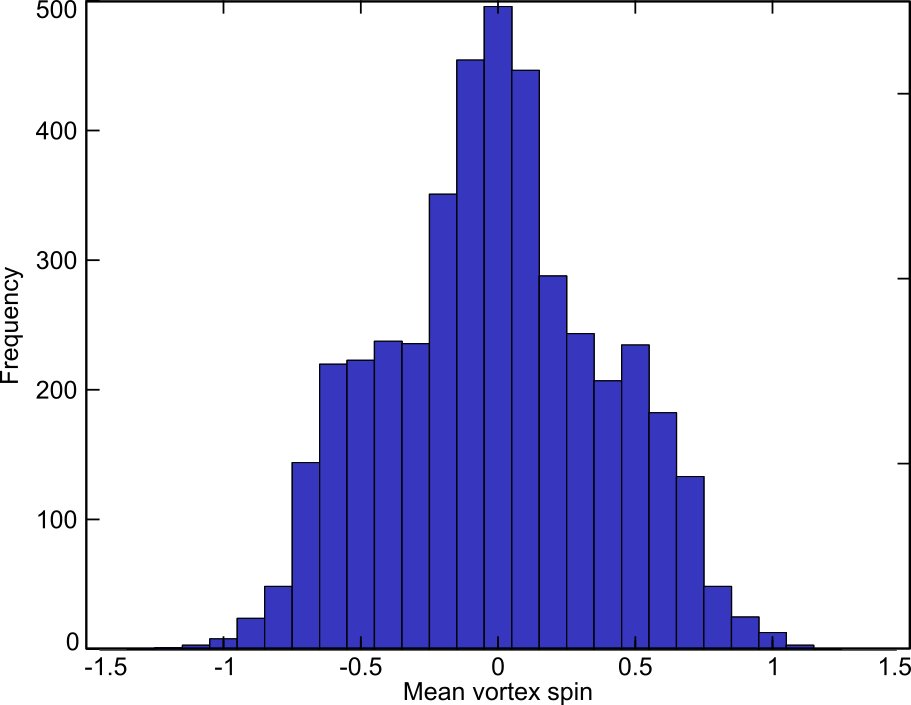}
	\caption{
\textbf{Vortices have no intrinsic rotation-sense bias.}
Histogram of time-averaged vortex spin of each cavity $\langle V_i(t) \rangle_t$ across all square lattice experiments, exhibiting symmetry about zero spin.}
\end{figure}

\begin{figure}[h!]
	\centering
	\includegraphics[width=\textwidth]{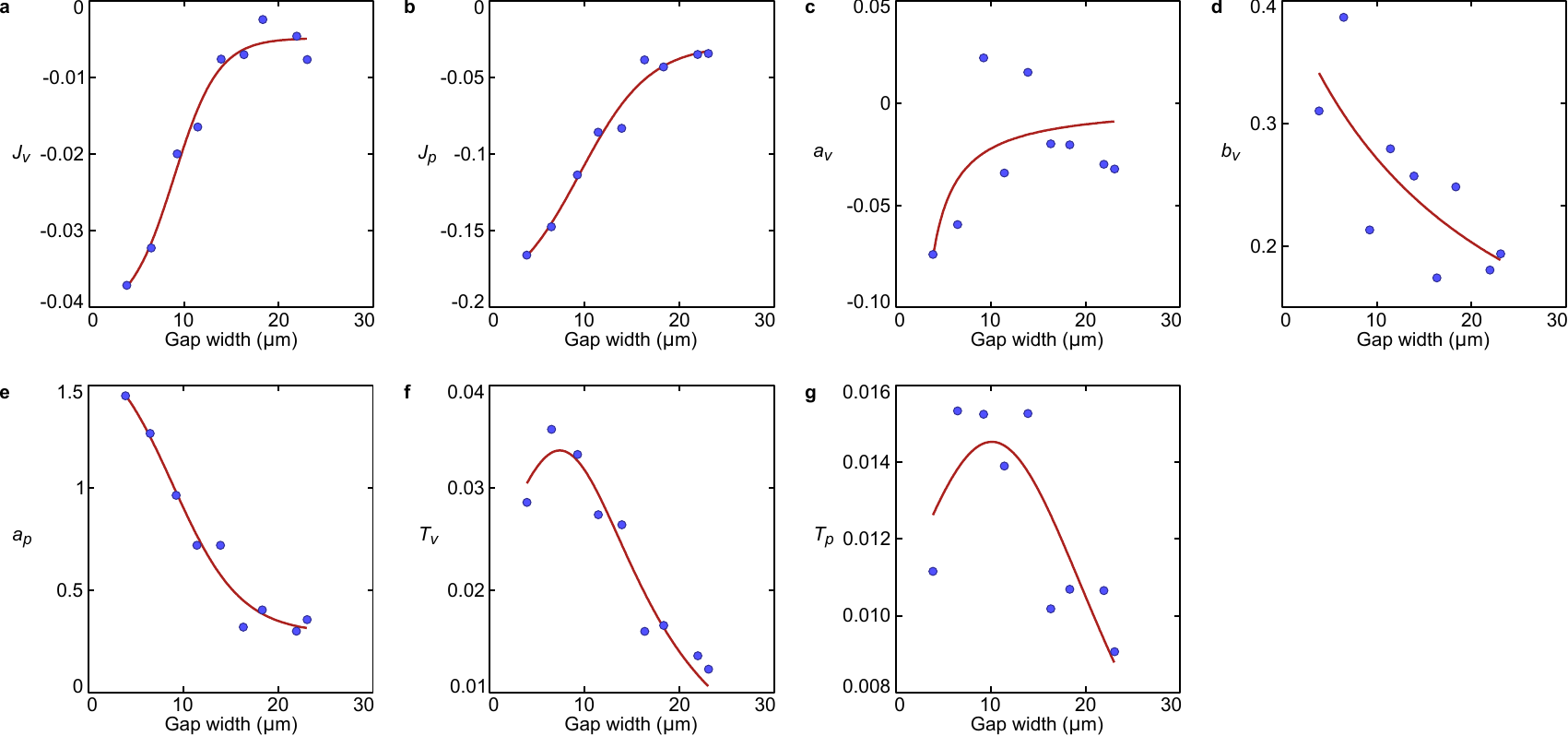}
	\caption{
\textbf{Parameters in the full model can be inferred using regression methods.}
Points are averages within non-overlapping $2.5\,\mu$m bins of parameters inferred for each experiment using linear regression on a discretization of Eq.~\eqref{e:SDE_model}, and lines are parametric best fits of selected functional forms to the points (Methods). }
\end{figure}

\begin{figure}[h!]
	\centering
	\includegraphics[width=89mm]{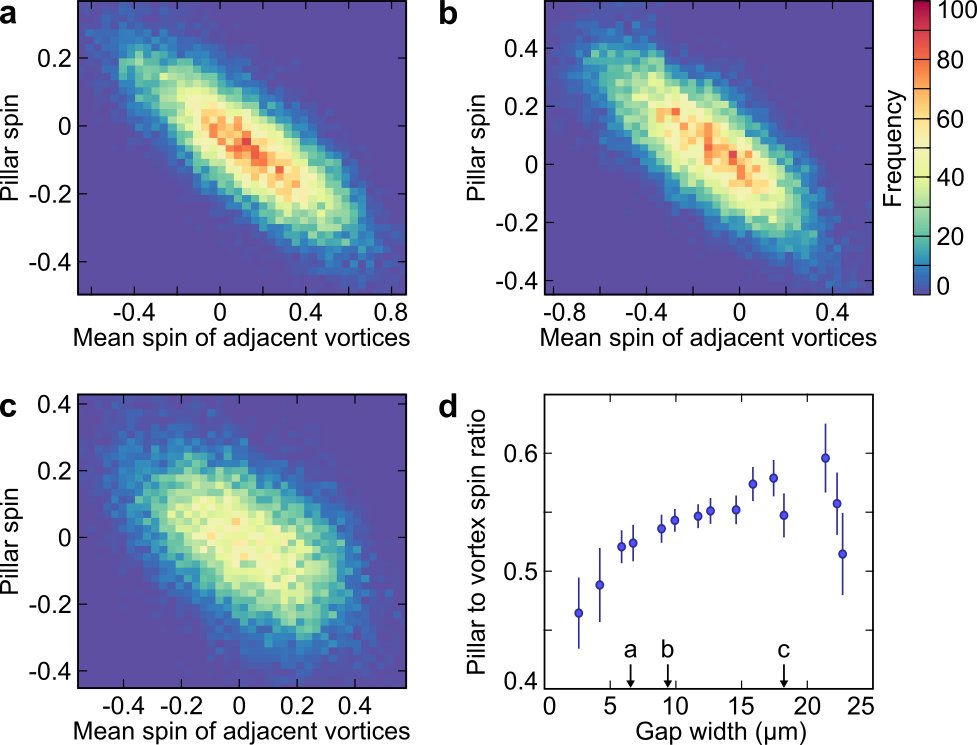}
	\caption{
\textbf{Pillar-spin distributions vary linearly with the average spin of adjacent vortices.}
\textbf{a}-\textbf{c}, Two-dimensional histogram of $(P_i, [P_i]_V)$ from three example movies, showing uniform spread about a line $P_i \propto [P_i]_V$. Gap widths $7$, $10$ and $18\,\um$, respectively.
\textbf{d}, The proportionality constant $\alpha$, where $P_i \approx -\alpha[P_i]_V$, depends weakly on gap width. Each point represents an average over $\geq 5$ movies in $3\,\um$ bins at $1.5\,\um$ intervals; vertical bars indicate standard errors (Methods).}
\end{figure}

\begin{figure}[h!]
	\centering
	\includegraphics[width=120mm]{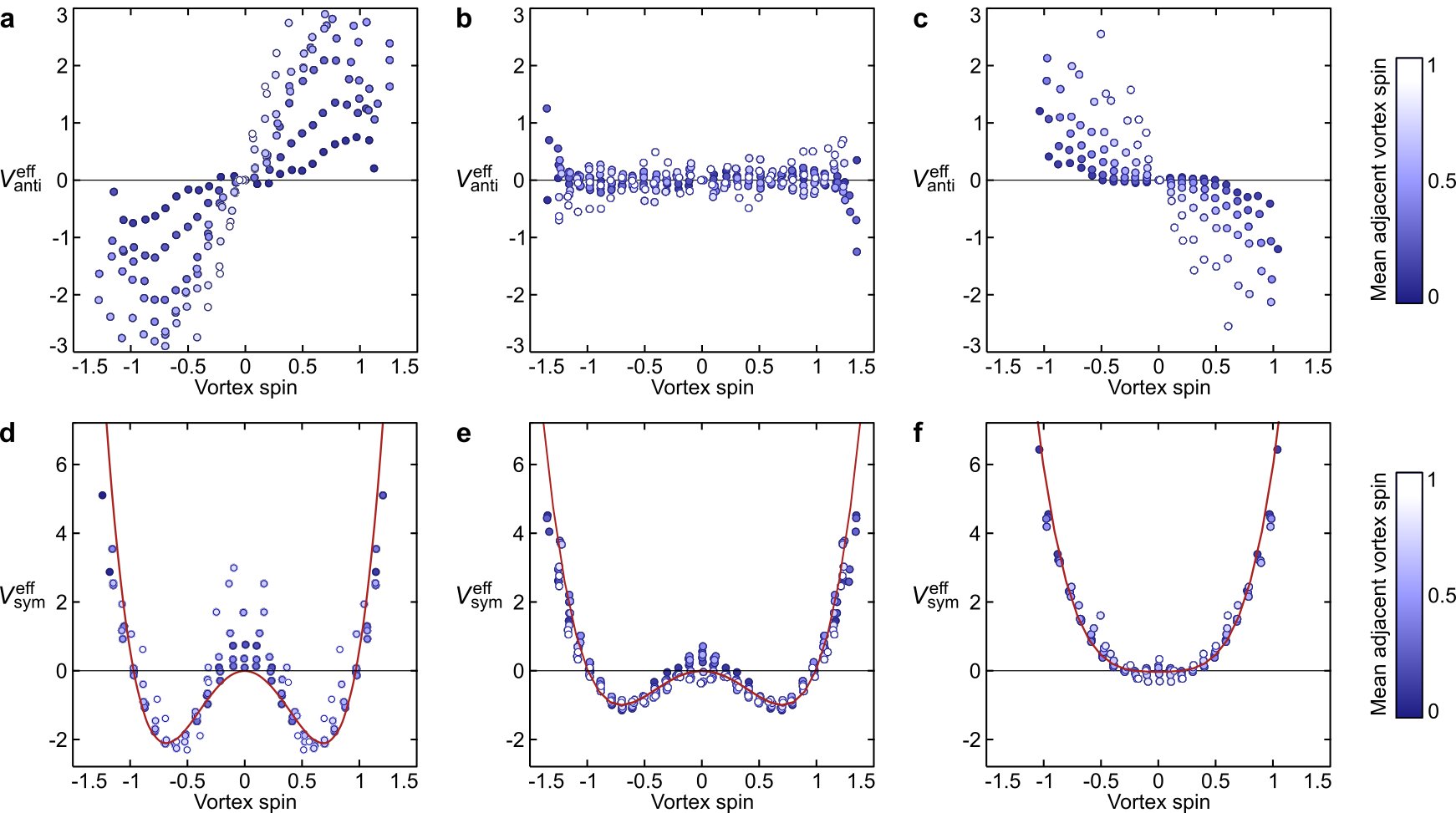}
	\caption{
\textbf{Parameters in the reduced model can be inferred by fitting effective single-spin potentials.} Reduced model parameters $\beta J$, $\beta a$ and $\beta b$ are estimated by fitting the antisymmetric and symmetric parts of the effective potential $\mathcal{V}^\mathrm{eff}(V_i \,|\, [V_i]_V)$ (Methods).  Data shown from three example movies of square lattices with gap widths $7\,\um$ (\textbf{a,d}), $10\,\um$ (\textbf{b,e}) and $18\,\um$ (\textbf{c,f}).
\textbf{a}--\textbf{c}, The antisymmetric part of the effective potential reveals the vortex--vortex coupling $\beta J$, spanning the range $J < 0$ (\textbf{a}), $J \approx 0$ (\textbf{b}) and $J > 0$ (\textbf{c}). Estimated $\mathcal{V}^\mathrm{eff}_\mathrm{anti}$ (points), coloured by mean adjacent spin $[V_i]_V$.
\textbf{d}--\textbf{f}, The symmetric part of the effective potential reveals the non-interacting single-spin potential, which flattens with increasing gap width. Estimated $\mathcal{V}^\mathrm{eff}_\mathrm{sym}$ (points) coloured by mean adjacent spin $[V_i]_V$, with fitted single-spin potentials (lines).}
\end{figure}

\begin{figure}[h!]
	\centering
	\includegraphics[width=89mm]{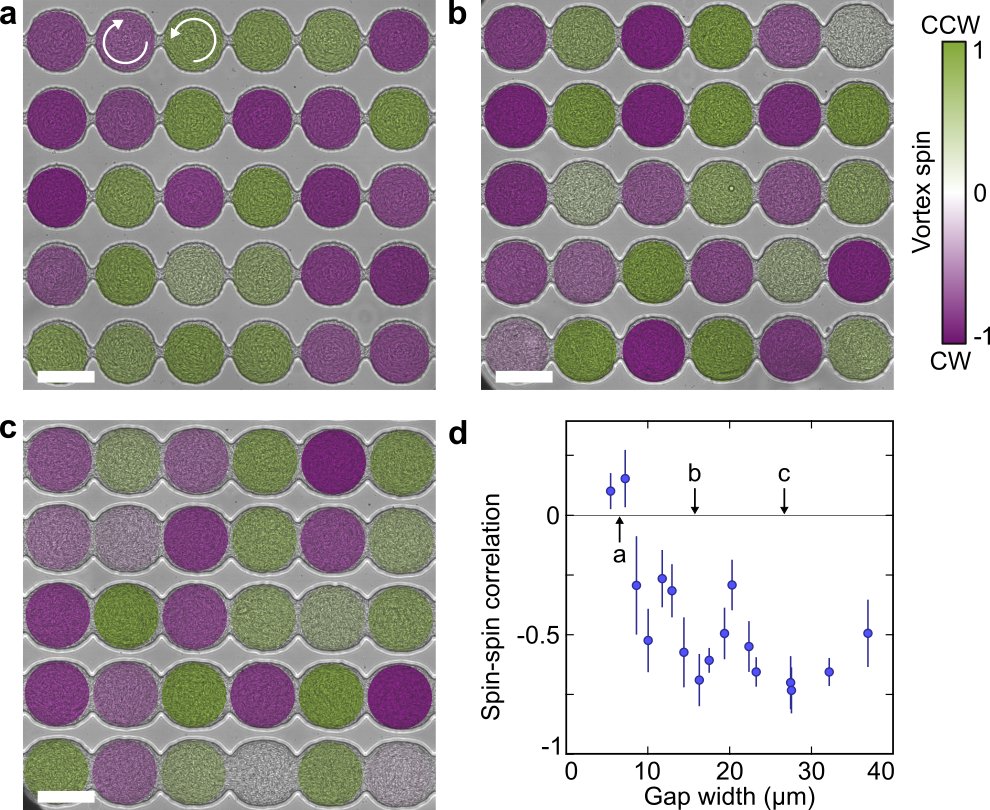}
	\caption{
\textbf{One-dimensional lattices adopt antiferromagnetic states.}
\textbf{a}, At the smallest gap widths, vortices interact weakly resulting in strong but randomly oriented circulation. Gap width $7\,\um$.
\textbf{b},\textbf{c}, Intermediate and large gaps show strong antiferromagnetic order. Gap widths $16$ and $28\,\um$. False colour in \textbf{a}--\textbf{c} denotes measured vortex spin. Scale bar: $50\,\um$.
\textbf{d}, The spin--spin correlation $\chi$ shows the antiferromagnetic state to be largely favoured in lines of cavities. Each point represents an average over $\geq 5$ movies in $3\,\um$ bins at $1.5\,\um$ intervals; vertical bars indicate standard errors (Methods).}
\end{figure}

\end{document}